\documentclass[10pt,conference]{IEEEtran}
\IEEEoverridecommandlockouts
\usepackage{epsfig,rotating,setspace,latexsym,amsmath,epsf,amssymb,amsfonts,bm,bbm,theorem,subfigure,epstopdf,cite,authblk,algcompatible,algorithm,caption}

\algrenewcommand\algorithmicforall{\textbf{foreach}}
\algrenewcommand\algorithmicindent{.8em}

\usepackage{color,nicefrac}
\usepackage{xcolor}
\usepackage{mathtools}

\newtheorem{theorem}{Theorem}

\newtheorem{problem}{Problem}

\IEEEoverridecommandlockouts
\allowdisplaybreaks
\usepackage{multirow}
\usepackage{soul}

\definecolor{Emerald}{rgb}{0.31, 0.78, 0.47}

\definecolor{RoyalBlue}{cmyk}{1, 0.50, 0, 0}

\definecolor{green1}{rgb}{0.2,0.7,0.2}

\begin{document}

\title{Fully Decentralized Task Offloading in Multi-Access Edge Computing Systems\thanks{Research of SA, MAZ, MB and TB was supported in part by the ARO MURI Grant AG285 and in part by the AFOSR Grant FA9550-24-1-0152. Emails: \texttt{\{sa57,mazaman2,basar1\}@illinois.edu}, \texttt{bastopcu@bilkent.edu.tr}, \texttt{ulukus@umd.edu}.
}}

\author[1]{Shubham Aggarwal}
\author[1]{Muhammad Aneeq uz Zaman}
\author[2]{Melih Bastopcu}
\author[3]{Sennur Ulukus}
\author[1]{Tamer Ba{\c s}ar}

\affil[1]{\normalsize Coordinated Science Laboratory, University of Illinois Urbana Champaign, Urbana, IL-61801}
\affil[2]{\normalsize Department of Electrical and Electronics Engineering, Bilkent University, Ankara, Türkiye, 06800}
\affil[3]{\normalsize Department of Electrical and Computer Engineering, University of Maryland, College Park, MD 20742}

\maketitle

\begin{abstract}
We consider the problem of task offloading in multi-access edge computing (MEC) systems constituting $N$ devices assisted by an edge server (ES), where the devices can split task execution between a local processor and the ES. Since the local task execution and communication with the ES both consume power, each device must judiciously choose between the two. We model the problem as a large population non-cooperative game among the $N$ devices. Since computation of an equilibrium policy in this large-device scenario can be extremely difficult, and can incur significant communication overhead, we employ the mean-field game framework to compute fully decentralized low complexity solutions for each device. By leveraging the novel age of information (AoI) metric, we invoke techniques from stochastic hybrid systems (SHS) theory to study the tradeoffs between increasing information freshness and reducing power consumption. In numerical results, we verify that a higher load at the ES may lead devices to push the tasks to the ES less often.   
\end{abstract}
\vspace{-0.5em}
\section{Introduction}
\vspace{-0.3em}
The multi-access edge computing (MEC) technology has recently attracted wide attention as a promising solution to improve computing capabilities, especially in the resource-limited dense networks of internet-of-things (IoT) devices \cite{mao2017survey,azmy2024incentive}. The MEC architecture leverages advances in wireless communication and mobile computing paradigms to allow for offloading task execution to the edge of the network. Edge computing is anticipated to play a crucial role in time-critical applications such as vehicle positioning in autonomous driving, task assignment problems in warehouses, and remote surgery systems \cite{hua2018energy, muhammad2021minimizing}, which form some of the major use-cases of future 6G networks.

\begin{figure}[t]
    \centerline{\includegraphics[width=0.74\columnwidth]{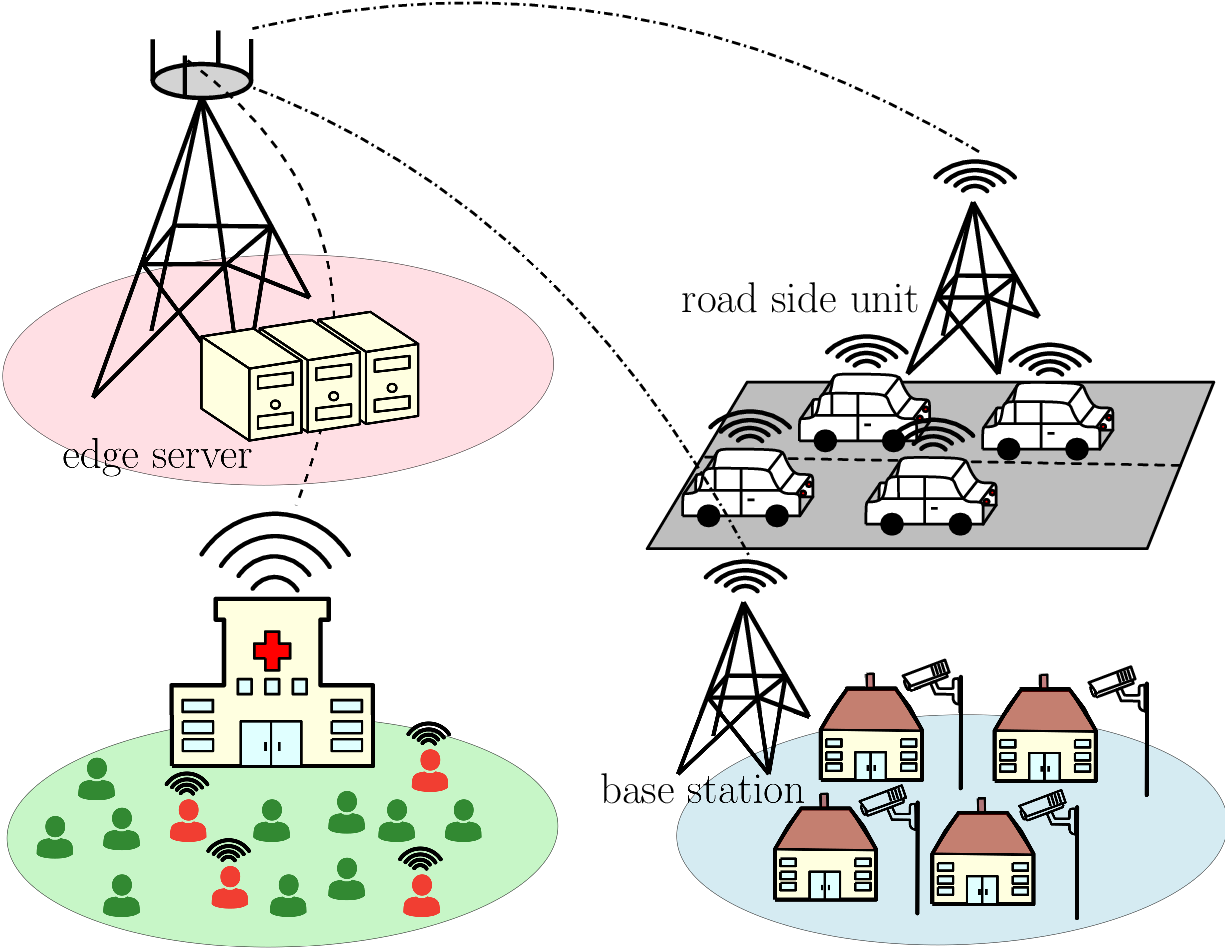}}\vspace{-0.25cm}
    \caption{\small{A MEC system model consisting of an edge server (ES) and various applications (medical, vehicular and home surveillance examples are shown in the figure) that utilize the ES for timely computation simultaneously.}}
    \label{Fig:system_model}
    \vspace{-0.5cm}
\end{figure}

In this work, we aim to: 1) accelerate task execution in MEC-based applications (hence, improve their situational awareness) by employing the novel age of information (AoI) metric \cite{yates2020age}, and 2) provide a low-complexity decentralized computation offloading algorithm for IoT devices in densely populated environments. Precisely, to reduce the high time complexity posed by centralized modeling schemes \cite{zhou2020partial}, we model the computation offloading problem in MEC systems comprising $N$ devices and an edge server (ES) using the framework of \textit{non-cooperative game theory}. An example of such a MEC system is shown in Fig.~\ref{Fig:system_model}, where in various applications (such as medical, vehicular and home surveillance examples as shown in the figure), devices offload a part of their computation to an ES. To entail tractable equilibrium policy computations, we employ the paradigm of mean-field games (MFGs) \cite{huang2004uplink, huang2007large, lasry2007mean, wang2014mean, aggarwal2023large, aggarwal2023weighted}, to compute approximate Nash equilibrium policies which ensure optimal division of task processing  between local processor and the ES.

\textit{Related Work:} Earlier work on the subject of computation offloading in MEC systems have focused on minimizing energy consumption, studying power-delay tradeoffs and server-device load balancing problems \cite{mao2016power, wang2017joint, mach2017mobile, mao2017survey} through the lens of centralized resource allocation involving multiple users. The latter problem is then solved using the Lyapunov optimization technique \cite{neely2022stochastic, jia2022lyapunov} to provide feasible solutions. Such algorithms can face high time complexity with limited scalability, especially in systems with large user populations \cite{zhou2020partial}. To address this issue, recent works have focused on game theoretic formulations \cite{messous2017computation,li2018game,wang2019game,zhou2020partial} to allow for decentralized decision making to achieve competing objectives of optimizing queuing theory-driven metrics of performance. Alternatively, freshness sensitive applications have employed the novel age-of-information based objective to maintain timeliness constraints at the end-user
\cite{ning2020mobile,sathyavageeswaran2024timely}. The issue of scalability, however, even within the game theoretic framework, still remains largely open.

Thus, the contribution and distinctiveness of our work are outlined as follows. We model the MEC problem using a game theoretic framework. To address the issues posed by scalability, particularly in ultra-dense user scenarios, we employ the novel MFG paradigm to compute completely decentralized offloading strategies for the end-users. Such a technique (1) alleviates the problem of Nash equilibrium computation in large-user games while allowing for tractable policy design, and (2) reduces the significant communication overhead which is required as part of the Nash equilibrium achievement. We further provide a low dimensional algorithm to compute approximate (local) Nash equilibria for the game problem. In the process, we also obtain closed-form AoI expressions for a system of series-parallel heterogeneous servers, which is a novel result of independent interest. The above features of scalability, decentralization and new quality-of-service metrics for maintaining freshness are significantly desirable in 5G advanced and future 6G scenarios where resource management becomes crucial, especially in ultra-dense user networks. 

\textbf{Notations:} $[N]:=\{1,\ldots,N\}$ denotes the set of users. We use the shorthand $\text{exp}(\lambda)$ to denote an exponential random variable with rate $\lambda$. For a policy vector $a = [a_1, \cdots, a_N]$, $a_{-i}$ denotes the policy vector of all users other than user $i$.

\section{$N$-User MEC Game Problem}
\begin{figure}[h]
    \centering
    \includegraphics[width=0.75\columnwidth]{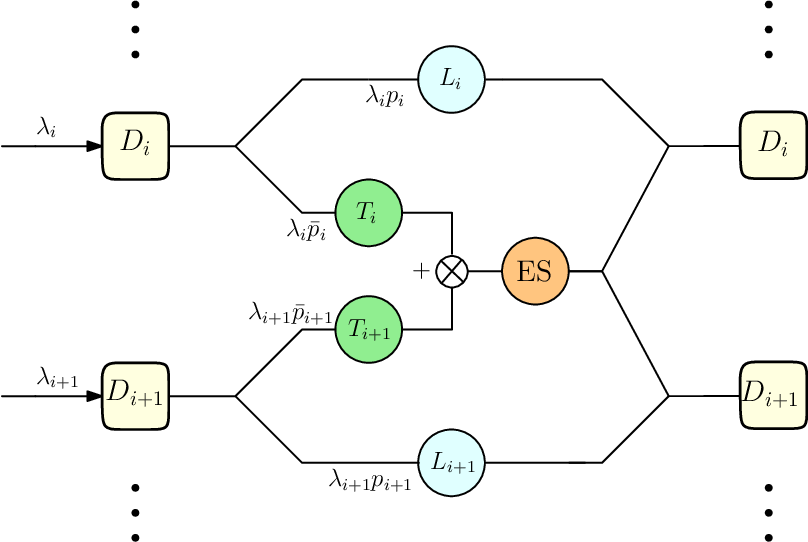}
    \caption{\small{Information flow schematic in a MEC system, For device $D_i$, $L_i$ and $T_i$ denote its local processor, and the transmitter, respectively.}}
    \label{fig:N_user}
    \vspace{-0.3cm}
\end{figure}

Consider the system in Fig.~\ref{fig:N_user} comprising $N$ devices which need to execute their respective incoming tasks. To handle heterogeneity among devices (for instance, with respect to device parameters or incoming task rates of the service parameters), we associate with each device a type parameter $\phi$ which belongs to a finite type set $\Phi$, and is sampled according to a probability distribution $\mathbb{P}_N(\phi)$, which appropriately accounts for the heterogeneity among the devices . To assist the devices with task execution, an ES is available. Thus, each device $D_i$ has two options for executing each incoming task: it can either serve it directly using its local processor ($L_i$) or it can offload it to the ES using its device transmitter ($T_i$), as shown in Fig. \ref{fig:N_user}. The inter-arrival times of tasks arriving at the $i$th device $D_i$ are distributed as a $\text{exp}(\lambda_i)$ random variable (r.v.) for all $i \in [N]$. If device $i$ decides to carry out the tasks on $L_i$, then it can operate the processor at a frequency $\mu_{1i} \leq f_{i,max}$. The service time of $L_i$ is distributed as an $\text{exp}(\mu_{1i})$ r.v. Accordingly, the \emph{processing power} used is $P_{\ell,i} = \eta \mu_{1i}^3$, where $\eta_i$ is a positive constant denoting the processor's effective capacitance \cite{mao2016power}.

On the other hand, if a device decides to offload the task to the ES, then it gets served sequentially by  $T_i$ to the ES and the ES uploads it back to the device after processing. The transmission rate of $T_i$ is modeled as an $\text{exp}(\mu_{2i})$ r.v. with $\mu_{2i}$ being the mean \emph{transmission power usage} and $\mu_{2i} \leq P_{i,max}$. The service time of task processing at the ES is modeled as an $\text{exp}(\mu_3^{(N)})$ r.v. where the superscript on $\mu_3^{(N)}$ (which is $\gg \mu_{1i}, \mu_{2i}$) denotes its dependence on the number of devices in the population. We employ the \textit{last-come-first-serve with preemption} (LCFS-P) discipline\footnote{The motivation behind using a preemption based discipline is two-fold: 1) it allows for efficient operation of systems with shared resources and selfish users (quantified using the price of anarchy or the price of stability metrics) as it has been observed in literature, see for instance, \cite{gai2016packet}, and 2) it allows for a manageable state space to compute average AoI expressions for a system with a hybrid connection of series-parallel servers. It would still be further interesting to theoretically analyze how our game theoretic based decentralized approach performs compared to a central optimization problem, and we leave this as an interesting future direction.} at all the servers ($L_i,T_i$, and the ES). Further, we assume that the downloading time of the processed task by the device is negligible (since they are usually low-bit sized commands such as collision warnings or real-time positions, for instance, in auto-driving scenarios).
% .\footnote{This is motivated by scenarios in autonomous vehicular systems or real-time monitoring systems where the uploaded tasks consist of high quality images or videos which take non-negligible transmission duration versus the processed tasks, which constitute low size commands (such as accident ahead, target's real-time position, etc.) which can be transmitted back to the IoT devices instantaneously. Also, the ES can be directly connected to the power source, and thus, it can use significantly higher transmission power. Since IoT devices have small batteries, their transmission times may not be negligible.} 

Since the effective service rates provided by $L_i$ and the series path of $T_i$ and the ES are heterogeneous, we employ the i.i.d.~Bernoulli distributed random variable with a mean $p_i$ to split the incoming Poisson process into two independent Poisson processes with respective means $\lambda_i p_i$ and $\lambda_i \Bar{p}_i$ where $\Bar{p}_i  =1-p_i$ (as in Fig.~\ref{fig:N_user}). Such a Bernoulli splitting has been widely employed in the literature in systems with heterogeneous parallel paths \cite{yates2018status}. Finally, we measure the freshness of processed information at the device using the average age-of-information (AoI) metric which is  defined as the time elapsed at the receiving end since the latest delivered information packet was generated at the source, and serves to quantify the freshness of information at the receiver.
% We note here that traditionally, queue length has been used to measure execution delay \cite{mao2016power}, which does not take into account the time criticality of the associated tasks. Since time responsiveness is one of the major concerns of this work, we employ the AoI metric which has been widely used to maintain timely updates of the tasks \cite{yates2020age}.

Thus, the objective of each device is two-fold: (1) To minimize the average AoI of the tasks, and (2) to minimize the power usage during local processing and transmission. Since this is a multi-objective optimization problem, in the sequel we use the scalarization approach \cite{boyd2004convex} to setup each device's problem. Let us define $\bm{\mu}_1:= [\mu_{11}, \cdots, \mu_{1N}]$, $\bm{\mu}_2:= [\mu_{21}, \cdots, \mu_{2N}]$ and $\bm{p}:= [p_1, \cdots, p_N]$. Then, the fraction of time that $L_i$ is busy can be computed as $t_{L_i} = \nicefrac{\lambda_i p_i}{(\lambda_i p_i + \mu_{2i})}$ and the fraction of time that $T_i$ is busy can be computed as $t_{T_i} = \nicefrac{\lambda_i \bar p_i}{(\lambda_i \bar p_i + \mu_{1i})}$. Consequently, each device $i \in [N]$ wishes to solve the following problem.
\begin{problem}[$N$--user game problem]\label{problem:N_user_game}
    \begin{align}  \min_{p_i,\mu_{1i},\mu_{2i}} & J_{N,i} (\bm{p},\bm{\mu}_1,\bm{\mu}_2) := t_{L_i}\mu_{1i} + t_{T_i}\eta \mu_{2i}^3 \nonumber \\
    & \hspace{3cm} + V \Delta^{(N)}_{i}\!(\bm{p},\bm{\mu}_1,\bm{\mu}_2) \nonumber\\
    \mbox{s.t.} \quad & \mu_{1i} \leq P_{i,max},~~ \mu_{2i} \leq f_{i,max} \label{prob1}
\end{align}
where $V>0$ is the importance weight given for freshness. 
\end{problem}

We also refer to the triple $(p_i,\mu_{1i},\mu_{2i})$ as the policy of device $i$. The problem in (\ref{prob1}) is a game problem due to the presence of other devices' policies in the cost optimization problem of the $i$th device. This requires each device to know the policy of the other devices to compute its own, which can incur a significant communication overhead, especially in a large user scenario. Thus, we will later employ the mean-field game framework to alleviate this issue and allow for tractable policy design. However, first, to completely formulate the above problem, we need to characterize the expression for the average AoI, $\Delta^{(N)}_i(\bm{p},\bm{\mu}_1,\bm{\mu}_2)$, which we will derive in the next section.

\section{Age of Information Calculation}
We calculate the AoI of the $i$th device using the SHS technique which utilizes tools from control theory and dynamical systems \cite{goebel2009hybrid, hespanha2006modelling, yates2018age} to handle systems involving both discrete and continuous states. For completeness, we briefly review the main concepts of the SHS method.

\subsection{Stochastic Hybrid Systems (SHS)}\label{Subsec:SHS}
\begin{figure}[h]
    \centering
    \includegraphics[width=0.60\columnwidth]{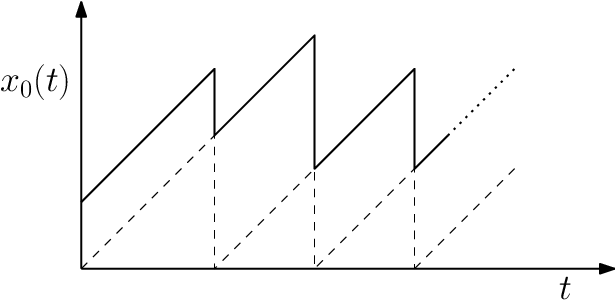}
    \caption{\small{Evolution of the AoI at the receiver.}}
    \label{fig:Sampe_AoI}
    \vspace{-0.3cm}
\end{figure}

The SHS method constitutes a state pair $(s(t),x(t)) \in S \times \mathbb{R}^{n+1}$ for all time $t \geq 0$, where $n+1$ denotes the number of servers including the device itself (with the device labelled as server 0). Further, $S$ is a finite set. The continuous state $x(t)$ evolves according to a stochastic differential equation $dx(t) = e(t,s,x)dt + g(t,s,x)dB(t)$, where $B(t)$ is a standard Brownian motion. Further, the discrete state $s(t)$ evolves according to a Markov chain from a state $s$ to a state $s'$ with transition intensity $q\delta_{s,s'}$, where $\delta_{s(t) = s'}=1$, if $s(t) = s'$, and 0 otherwise. At each transition, the jump in the continuous state is given as $x' = h(t,s,x)$. With the above description, AoI can be characterized as a special case of the SHS framework. A prototypical sample path of the AoI evolution is shown in Fig.~\ref{fig:Sampe_AoI}, which is a piecewise linear SHS with $e(t,s,x) = u_s,$ $ g(t,s,x) = 0,$ $u_s \in \{0,1\}$, and $h(t,s,x) = xA_s$, where $A_s \in \{0,1\}^{(n+1) \times (n+1)}$.

Following \cite{yates2018age}, we now define $\pi_{s'}(t):= \mathbb{P}(s(t) = s')$ as the probability of state $s(t) = s'$, and $v_{s'k}(t):= \mathbb{E}[x_k(t) \delta_{s(t) = s'}]$ to measure the correlation between the AoI process $x_k(t)$ in server $k$ with the state $s(t)$ at timestep $t$ with $v_s:= [v_{s0}, \cdots, v_{sn}]$. Further, let us denote the set of possible outgoing transitions from a particular state $s$ as $L_s:= \{\ell: s_{\ell} = s\}$ and the set of possible incoming transitions to a state $s'$ as $L'_{s'}:= \{\ell: s_\ell = s'\}$. Then, assuming that the finite state Markov chain (FS-MC) is ergodic, it has a unique steady state distribution $\Bar{\pi}:= [\bar{\pi}_1, \cdots, \bar{\pi}_m]$, which satisfies the conservation law,
\begin{subequations}\label{steady_state_prob}
\begin{small}
    \begin{align}
    \Bar{\pi}_s \sum_{\ell \in L_s}q^\ell & = \sum_{\ell' \in L'_s} q^{\ell'} \Bar{\pi}_{s_{\ell'}}, \quad \forall s \in S, \\
    \sum_{s \in S} \Bar{\pi}_s &= 1, \label{prob_sum}
\end{align}
\end{small}
\end{subequations}
where $m := |S|$. Consequently, we have the following result.

\begin{theorem}\!\cite[Thm.~4]{yates2018age}\label{Avg_Age_thm}
    Suppose $\Bar{\pi}$ is the state distribution of the FS-MC and there exists a stationary solution $\Bar{v} := [\Bar{v}_1,\cdots, \Bar{v}_m]$ of the conditional distribution $v_{(\cdot)}(t)$ satisfying,
    \begin{align}\label{cond_prob_eqn}
        \Bar{v}_s \sum_{\ell \in L_s}q^\ell = u_s \Bar{\pi}_s + \sum_{\ell' \in L'_s} q^{\ell'} \Bar{v}_{s_{\ell'}}A_{\ell'}.
    \end{align}
    Then, the average AoI is given by $\Delta:= \sum_{s \in S}\bar{v}_{s0}$.
\end{theorem}

Next, we will use the above result to compute an approximate expression for ${\Delta}^{(N)}_i(\bm{p},\bm{\mu}_1,\bm{\mu}_2)$ in the next subsection.

\subsection{Average AoI for the $i$th Device}
\begin{figure}[h]
    \centering
    \includegraphics[width=0.8\columnwidth]{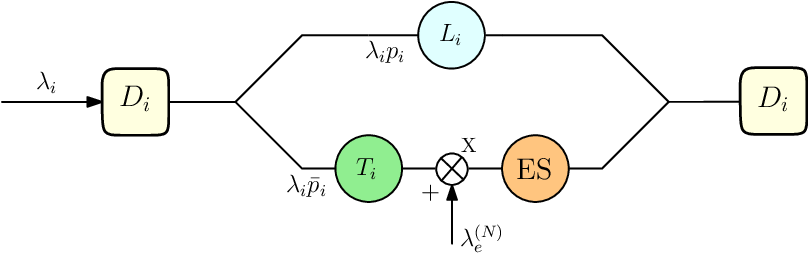}
    \caption{\small{Task flow for device $i$: $D_i$, $L_i$, $T_i$ denote the $i$th device itself, its local processor, and its transmitter, respectively.}}
    \label{fig:generic_user}
    \vspace{-0.2cm}
\end{figure}

Let us consider the task flow from the perspective of the $ith$ device, as shown in Fig.~\ref{fig:generic_user}, where the interference from the other devices is denoted by the adder (marked as X) preceding the ES, which receives packets according to an exogenous process with a combined rate of $\lambda^{(N)}_e$. We first observe that this process may not obey a Poisson distribution, which makes it challenging to compute the exact expression for ${\Delta}^{(N)}_i(\bm{p},\bm{\mu}_1,\bm{\mu}_2)$. Additionally, the same prevents us from utilizing the fake update approach for all servers as proposed in \cite{yates2018age}, which is very effective in performing a computationally reduced dimensional analysis. Thus, to facilitate tractability, we take $\lambda_i$'s to be large, in which case, the aforementioned distribution can be closely approximated by a $\text{exp}(\lambda_e^{(N)})$ distribution \cite{yates2018age}, where we define $\lambda_e^{(N)}:= \sum_{j=1, j \ne i}^N \frac{\lambda_j\bar{p}_j \mu_{1j}}{\lambda_j\bar{p}_j + \mu_{1j}}$.

Next, we formulate the state space and transition functions of the FS-MC. Henceforth, we refer to device $i$'s packets as those of class 1 and exogenous packets as those of `class 2'. 

\begin{table}[h!]
\vspace{-2mm}
\centering
\begin{tabular}{ |c|c|c|c| } 
\hline
state & server 1 ($T_i$) & server 2 $(L_i)$ & server 3 (ES) \\
\hline
$s_1$ & freshest & $2^{nd}$ freshest & oldest \\ \hline
$s_2$ & freshest & oldest & $2^{nd}$ freshest \\ \hline
$s_3$ & $2^{nd}$ freshest & freshest & oldest \\ \hline
$s_4$ & \emph{no packet} & freshest & $2^{nd}$ freshest  \\ \hline
$s_5$ & \emph{no packet} & $2^{nd}$freshest & freshest \\ \hline
$s_6$ & \emph{no packet} & freshest & class 2 \\ \hline
$s_7$ & freshest & $2^{nd}$ freshest & class 2 \\ \hline
$s_8$ & $2^{nd}$ freshest & freshest & class 2 \\
\hline
\end{tabular}
% \vspace{0.1cm}
\caption{\small{State dictionary for the finite FS-MC.}}
\label{table:states}
\vspace{-0.5cm}
\end{table}

The state space $S$ comprises 8 states which keep track of the server holding the freshest and  second freshest packets, and the oldest packet of class 1, and the server holding a packet of class 2. Detailed descriptions are provided in Table~\ref{table:states}.

    \begin{figure*}%
        \vspace{.5mm}
    \end{figure*}
\begin{table}[h]
\centering
% \vspace{-2mm}
\small
\begin{tabular}{|c|c|c|c|c|} 
\hline
$s$ & $q$ & $s'$ & $x' = xA_s$ & $\bar v_sA_s$\\
\hline
\multirow{4}{2em}{$s_1$} & $\lambda p$ & $s_3$ & $[x_0 ~x_1 ~0 ~x_3]$ & $[\bar{v}_{10}~\bar{v}_{11}~0~\bar{v}_{13}]$\\
& $\lambda\Bar{p}$ & $s_1$ & $[x_0 ~0 ~x_2 ~x_3]$ & $[\bar{v}_{10}~0~\bar{v}_{12}~\bar{v}_{13}]$ \\
& $\lambda_e^{(N)}$ & $s_7$ & $[x_0 ~x_1 ~x_2 ~x_0]$ & $[\bar{v}_{10}~\bar{v}_{11}~\bar{v}_{12}~\bar{v}_{10}]$ \\
& $\mu_1$ & $s_5$ & $[x_0 ~0 ~x_2 ~x_1]$ & $[\bar{v}_{10}~0~\bar{v}_{12}~\bar{v}_{11}]$ \\
& $\mu_2$ & $s_1$ & $[x_2 ~x_1 ~x_2 ~x_2]$ & $[\bar{v}_{12}~\bar{v}_{11}~\bar{v}_{12}~\bar{v}_{12}]$ \\
& $\mu_3^{(N)}$ & $s_1$ & $[x_3 ~x_1 ~x_2 ~x_3]$ & $[\bar{v}_{13}~\bar{v}_{11}~\bar{v}_{12}~\bar{v}_{13}]$ \\ \hline
\multirow{4}{2em}{$s_2$} & $\lambda p$ & $s_3$ & $[x_0 ~x_1 ~0 ~x_3]$ & $[\bar{v}_{20}~\bar{v}_{21}~0~\bar{v}_{23}]$ \\
& $\lambda \Bar{p}$ & $s_2$ & $[x_0 ~0 ~x_2 ~x_3]$ & $[\bar{v}_{20}~0~\bar{v}_{22}~\bar{v}_{23}]$ \\
& $\lambda_e^{(N)}$ & $s_7$ & $[x_0 ~x_1 ~x_2 ~x_0]$ & $[\bar{v}_{20}~\bar{v}_{21}~\bar{v}_{22}~\bar{v}_{20}]$ \\
& $\mu_1$ & $s_5$ & $[x_0 ~0 ~x_2 ~x_1]$ & $[\bar{v}_{20}~0~\bar{v}_{22}~\bar{v}_{21}]$ \\
& $\mu_2$ & $s_2$ & $[x_2 ~x_1 ~x_2 ~x_3]$ & $[\bar{v}_{22}~\bar{v}_{21}~\bar{v}_{22}~\bar{v}_{23}]$ \\
& $\mu_3^{(N)}$ & $s_2$ & $[x_3 ~x_1 ~x_3 ~x_3]$ & $[\bar{v}_{23}~\bar{v}_{21}~\bar{v}_{23}~\bar{v}_{23}]$ \\ \hline
\multirow{4}{2em}{$s_3$} & $\lambda p$ & $s_3$ & $[x_0 ~x_1 ~0 ~x_3]$ & $[\bar{v}_{30}~\bar{v}_{31}~0~\bar{v}_{33}]$ \\
& $\lambda \Bar{p}$ & $s_1$ & $[x_0 ~0 ~x_2 ~x_3]$ & $[\bar{v}_{30}~0~\bar{v}_{32}~\bar{v}_{33}]$ \\
& $\lambda_e^{(N)}$ & $s_8$ & $[x_0 ~x_1 ~x_2 ~x_0]$ & $[\bar{v}_{30}~\bar{v}_{31}~\bar{v}_{32}~\bar{v}_{30}]$ \\
& $\mu_1$ & $s_4$ & $[x_0 ~0 ~x_2 ~x_1]$ & $[\bar{v}_{30}~0~\bar{v}_{32}~\bar{v}_{31}]$ \\
& $\mu_2$ & $s_3$ & $[x_2 ~x_2 ~x_2 ~x_2]$ & $[\bar{v}_{32}~\bar{v}_{32}~\bar{v}_{32}~\bar{v}_{32}]$ \\
& $\mu_3^{(N)}$ & $s_3$ & $[x_3 ~x_1 ~x_2 ~x_3]$ & $[\bar{v}_{33}~\bar{v}_{31}~\bar{v}_{32}~\bar{v}_{33}]$ \\ \hline
\multirow{4}{2em}{$s_4$} & $\lambda p$ & $s_4$ & $[x_0 ~0 ~0 ~x_3]$ & $[\bar{v}_{40}~0~0~\bar{v}_{43}]$ \\
& $\lambda\Bar{p}$ & $s_1$ & $[x_0 ~0 ~x_2 ~x_3]$ & $[\bar{v}_{40}~0~\bar{v}_{42}~\bar{v}_{43}]$ \\
& $\lambda_e^{(N)}$ & $s_6$ & $[x_0 ~0 ~x_2 ~x_0]$ & $[\bar{v}_{40}~0~\bar{v}_{42}~\bar{v}_{40}]$ \\
& $\mu_2$ & $s_4$ & $[x_2 ~0 ~x_2 ~x_2]$ & $[\bar{v}_{42}~0~\bar{v}_{42}~\bar{v}_{42}]$ \\
& $\mu_3^{(N)}$ & $s_4$ & $[x_3 ~0 ~x_2 ~x_3]$ & $[\bar{v}_{43}~0~\bar{v}_{42}~\bar{v}_{43}]$ \\ \hline
\multirow{4}{2em}{$s_5$} & $\lambda p$ & $s_4$ & $[x_0 ~0 ~0 ~x_3]$ & $[\bar{v}_{50}~0~0~\bar{v}_{53}]$ \\
& $\lambda \Bar{p}$ & $s_2$ & $[x_0 ~0 ~x_2 ~x_3]$ & $[\bar{v}_{50}~0~\bar{v}_{52}~\bar{v}_{53}]$ \\
& $\lambda_e^{(N)}$ & $s_6$ & $[x_0 ~0 ~x_2 ~x_0]$ & $[\bar{v}_{50}~0~\bar{v}_{52}~\bar{v}_{50}]$ \\
& $\mu_2$ & $s_5$ & $[x_2 ~0 ~x_2 ~x_3]$ & $[\bar{v}_{52}~0~\bar{v}_{52}~\bar{v}_{53}]$ \\
& $\mu_3^{(N)}$ & $s_5$ & $[x_3 ~0 ~x_3 ~x_3]$ & $[\bar{v}_{53}~0~\bar{v}_{53}~\bar{v}_{53}]$ \\ \hline
\multirow{4}{2em}{$s_6$} & $\lambda p$ & $s_6$ & $[x_0 ~0 ~0 ~x_3]$ & $[\bar{v}_{60}~0~0~\bar{v}_{63}]$ \\
& $\lambda \Bar{p}$ & $s_7$ & $[x_0 ~0 ~x_2 ~x_3]$ & $[\bar{v}_{60}~0~\bar{v}_{62}~\bar{v}_{63}]$ \\
& $\lambda_e^{(N)}$ & $s_6$ & $[x_0 ~0 ~x_2 ~x_0]$ & $[\bar{v}_{60}~0~\bar{v}_{62}~\bar{v}_{60}]$ \\
& $\mu_2$ & $s_6$ & $[x_2 ~0 ~x_2 ~x_2]$ & $[\bar{v}_{62}~0~\bar{v}_{62}~\bar{v}_{62}]$ \\
& $\mu_3^{(N)}$ & $s_6$ & $[x_3 ~0 ~x_2 ~x_3]$ & $[\bar{v}_{63}~0~\bar{v}_{62}~\bar{v}_{63}]$ \\ \hline
\multirow{4}{2em}{$s_7$} & $\lambda p$ & $s_8$ & $[x_0 ~x_1 ~0 ~x_3]$ & $[\bar{v}_{70}~\bar{v}_{71}~0~\bar{v}_{73}]$ \\
& $\lambda \Bar{p}$ & $s_7$ & $[x_0 ~0 ~x_2 ~x_3]$ & $[\bar{v}_{70}~0~\bar{v}_{72}~\bar{v}_{73}]$ \\
& $\lambda_e^{(N)}$ & $s_7$ & $[x_0 ~x_1 ~x_2 ~x_0]$ & $[\bar{v}_{70}~\bar{v}_{71}~\bar{v}_{72}~\bar{v}_{70}]$ \\
& $\mu_1$ & $s_5$ & $[x_0 ~0 ~x_2 ~x_1]$ & $[\bar{v}_{70}~0~\bar{v}_{72}~\bar{v}_{71}]$ \\
& $\mu_2$ & $s_7$ & $[x_2 ~x_1 ~x_2 ~x_2]$ & $[\bar{v}_{72}~\bar{v}_{71}~\bar{v}_{72}~\bar{v}_{72}]$ \\
& $\mu_3^{(N)}$ & $s_7$ & $[x_3 ~x_1 ~x_2 ~x_3]$ & $[\bar{v}_{73}~\bar{v}_{71}~\bar{v}_{72}~\bar{v}_{73}]$ \\ \hline
\multirow{4}{2em}{$s_8$} & $\lambda p$ & $s_8$ & $[x_0 ~x_1 ~0 ~x_3]$ & $[\bar{v}_{80}~\bar{v}_{81}~0~\bar{v}_{83}]$ \\
& $\lambda \Bar{p}$ & $s_7$ & $[x_0 ~0 ~x_2 ~x_3]$ & $[\bar{v}_{80}~0~\bar{v}_{82}~\bar{v}_{83}]$ \\
& $\lambda_e^{(N)}$ & $s_8$ & $[x_0 ~x_1 ~x_2 ~x_0]$ & $[\bar{v}_{80}~\bar{v}_{81}~\bar{v}_{82}~\bar{v}_{80}]$ \\
& $\mu_1$ & $s_4$ & $[x_0 ~0 ~x_2 ~x_1]$ & $[\bar{v}_{80}~0~\bar{v}_{82}~\bar{v}_{81}]$ \\
& $\mu_2$ & $s_8$ & $[x_2 ~x_2 ~x_2 ~x_2]$ & $[\bar{v}_{82}~\bar{v}_{82}~\bar{v}_{82}~\bar{v}_{82}]$ \\
& $\mu_3^{(N)}$ & $s_8$ & $[x_3 ~x_1 ~x_2 ~x_3]$ & $[\bar{v}_{83}~\bar{v}_{81}~\bar{v}_{82}~\bar{v}_{83}]$ \\ \hline
\end{tabular}
% \vspace{0.1cm}
\caption{\small{State transitions of the FS-MC and associated AoI jumps}}
\label{table:transitions}
\vspace{-0.7cm}
\end{table}

Next, in Table~\ref{table:transitions}, we list the possible transitions in the FS-MC and the corresponding AoI vector $x'(t):= [x'_0(t) ~x'_1(t)~x'_2(t)~x'_3(t)]$, where $x'_0(t),$ $x'_1(t),$ $x'_2(t)$, and $x'_3(t)$ denote the AoI at the $i$th device, the local processor, the transmitter, and the ES, respectively, after transition to $s'$. For example, the very first row can be read as follows: The system in Fig. \ref{fig:generic_user} transits from state $s_1$ to state $s_3$ when a new task arrives at $L_i$ with the corresponding AoI vector jumping to $x' = [x_0~x_1~0~x_3]$ and the conditional probability vector to $[\bar v_{10} ~\bar v_{11}~ 0 ~\bar v_{13}]$. Note that henceforth we forego the subscript index $i$ for brevity.

Finally, we without loss of generality, we can assume that all servers \emph{which do not precede a node of packet arrival} are busy all the time, i.e., whenever a packet leaves a server, a fake packet with the same type and AoI as the departing one starts processing. It is essential to take care of the emphasized statement, since in our case server 1 precedes the point of arrival of exogeneous packets. Thus, the SHS model should take into account whether it is idling or is busy, and hence, we \textit{cannot} run a fake update at this server. Consequently, we have that $u_s = [1~1~1~1]$ for $s=s_1,s_2,s_3,s_7,s_8,$ and $u_s:= \hat{u}_s = [1~0~1~1]$ for $s=s_4,s_5,s_6$.

Let us define $a:= \lambda + \lambda_e^{(N)} + \mu_1 + \mu_2 + \mu_3^{(N)}$ and $\hat{a} :=a-\mu_1$. Then, using \eqref{steady_state_prob}, $\Bar{\pi}$ satisfies \eqref{prob_sum} and the following equations,
\begin{subequations}\label{steady_state_prob_this_work}
\begin{small}
    \begin{align}\nonumber\\[-2.1em]
        a \Bar{\pi}_1 & = (\lambda\Bar{p} + \mu_2 + \mu_3^{(N)})\Bar{\pi}_1 + \lambda\Bar{p} (\Bar{\pi}_3 + \Bar{\pi}_4), \\ 
        a \Bar{\pi}_2 & = (\lambda\Bar{p} + \mu_2 + \mu_3^{(N)})\Bar{\pi}_2 + \lambda\Bar{p} \Bar{\pi}_5, \\ 
        a \Bar{\pi}_3 & = (\lambda p + \mu_2 + \mu_3^{(N)})\Bar{\pi}_3 + \lambda p (\Bar{\pi}_1 + \Bar{\pi}_2), \\ 
        \hat{a} \Bar{\pi}_4 & = (\lambda p + \mu_2 + \mu_3^{(N)})\Bar{\pi}_4 \!+\! \lambda p \Bar{\pi}_5 \!+\! \mu_1 (\Bar{\pi}_3 \!+\! \Bar{\pi}_8),\!\! \\ 
        \hat{a} \Bar{\pi}_5 & = (\mu_2 + \mu_3^{(N)})\Bar{\pi}_5 + \mu_1 (\Bar{\pi}_1 + \Bar{\pi}_2 + \Bar{\pi}_7), \\
        \hat{a} \Bar{\pi}_6 & = (\lambda p + \lambda_e^{(N)} + \mu_2 + \mu_3^{(N)})\Bar{\pi}_6 + \lambda_e^{(N)}(\Bar{\pi}_4 + \Bar{\pi}_5), \\
        a \Bar{\pi}_7 & = (\lambda\Bar{p} + \lambda_e^{(N)} + \mu_2 + \mu_3^{(N)})\Bar{\pi}_7 + \lambda_e^{(N)}(\Bar{\pi}_1 + \Bar{\pi}_2)\nonumber\\& ~~~+\! \lambda \Bar{p} (\Bar{\pi}_6 + \Bar{\pi}_8), \\ 
        a \Bar{\pi}_8 & = (\lambda p\! +\! \lambda_e^{(N)}\! \!+ \!\mu_2 \!+\! \mu_3^{(N)}\!)\Bar{\pi}_8 \!+ \!\lambda_e^{(N)} \!(\Bar{\pi}_3)\!+\! \lambda p \Bar{\pi}_7. \!\!
    \end{align}
    \end{small}
\end{subequations}

\begin{figure*}[h]
\begin{small}
\begin{subequations}\label{eqn_v_s}
    \begin{align}
        a\Bar{v}_1 & \!= \!u_s \Bar{\pi}_1 \!+\! \lambda \Bar{p}[\bar{v}_{10}~0~\bar{v}_{12}~\bar{v}_{13}] + \mu_2 [\bar{v}_{12}~\bar{v}_{11}~\bar{v}_{12}~\bar{v}_{12}] + \mu_3^{(N)} [\bar{v}_{13}~\bar{v}_{11}~\bar{v}_{12}~\bar{v}_{13}] + \lambda \Bar{p} [\bar{v}_{30}~0~\bar{v}_{32}~\bar{v}_{33}]  + \lambda \Bar{p} [\bar{v}_{40}~0~\bar{v}_{42}~\bar{v}_{43}] \\
        a\Bar{v}_2 & = u_s \Bar{\pi}_2 + \lambda \Bar{p}[\bar{v}_{20}~0~\bar{v}_{22}~\bar{v}_{23}] + \mu_2 [\bar{v}_{22}~\bar{v}_{21}~\bar{v}_{22}~\bar{v}_{23}] + \mu_3^{(N)} [\bar{v}_{23}~\bar{v}_{21}~\bar{v}_{23}~\bar{v}_{23}] + \lambda \Bar{p} [\bar{v}_{50}~0~\bar{v}_{52}~\bar{v}_{53}] \\
        a\Bar{v}_3 & \!= \!u_s \Bar{\pi}_3 + \lambda p [\bar{v}_{30}~\bar{v}_{31}~0~\bar{v}_{33}] \!+ \!\mu_2 [\bar{v}_{32}~\bar{v}_{32}~\bar{v}_{32}~\bar{v}_{32}] + \mu_3^{(N)} [\bar{v}_{33}~\bar{v}_{31}~\bar{v}_{32}~\bar{v}_{33}] + \lambda p [\bar{v}_{10}~\bar{v}_{11}~0~\bar{v}_{13}]  + \lambda p [\bar{v}_{20}~\bar{v}_{21}~0~\bar{v}_{23}] \\
        \hat{a}\Bar{v}_4 & \!=\! \hat{u}_s \Bar{\pi}_4 \!+\! \lambda p ([\bar{v}_{40}~0~0~\bar{v}_{43}] \!+\! [\bar{v}_{50}~0~0~\bar{v}_{53}]) \!+\! \mu_2 [\bar{v}_{42}~0~\bar{v}_{42}~\bar{v}_{42}] + \mu_3^{(N)} [\bar{v}_{43}~0~\bar{v}_{42}~\bar{v}_{43}] \!+\! \mu_1 ([\bar{v}_{30}~0~\bar{v}_{32}~\bar{v}_{31}]\!+\![\bar{v}_{80}~0~\bar{v}_{82}~\bar{v}_{81}])\\
        \hat{a}\Bar{v}_5 & = \hat{{u}}_s \Bar{\pi}_5 + \mu_1 [\bar{v}_{70}~0~\bar{v}_{72}~\bar{v}_{71}] + \mu_2 [\bar{v}_{52}~0~\bar{v}_{52}~\bar{v}_{53}] + \mu_3^{(N)} [\bar{v}_{53}~0~\bar{v}_{53}~\bar{v}_{53}] + \mu_1 [\bar{v}_{10}~0~\bar{v}_{12}~\bar{v}_{11}] + \mu_1 [\bar{v}_{20}~0~\bar{v}_{22}~\bar{v}_{21}] \\
        \hat{a}\Bar{v}_6 & \!=\! \hat{{u}}_s \Bar{\pi}_6 \!+\! \lambda p [\bar{v}_{60}~0~0~\bar{v}_{63}] \!+\! \lambda_e^{(N)} ([\bar{v}_{40}~0~\bar{v}_{42}~\bar{v}_{40}] \!+\! [\bar{v}_{50}~0~\bar{v}_{52}~\bar{v}_{50}] \!+\! [\bar{v}_{60}~0~\bar{v}_{62}~\bar{v}_{60}]) \!+\! \mu_2 [\bar{v}_{62}~0~\bar{v}_{62}~\bar{v}_{62}] \!+\! \mu_3^{(N)} [\bar{v}_{63}~0~\bar{v}_{62}~\bar{v}_{63}] \\
        a\Bar{v}_7 & = u_s \Bar{\pi}_7 + \lambda \Bar{p}[\bar{v}_{60}~0~\bar{v}_{62}~\bar{v}_{63}] + \lambda \Bar{p}[\bar{v}_{70}~0~\bar{v}_{72}~\bar{v}_{73}] + \lambda \Bar{p}[\bar{v}_{80}~0~\bar{v}_{82}~\bar{v}_{83}] + \lambda_e^{(N)} [\bar{v}_{10}~\bar{v}_{11}~\bar{v}_{12}~\bar{v}_{10}] + \lambda_e^{(N)} [\bar{v}_{20}~\bar{v}_{21}~\bar{v}_{22}~\bar{v}_{20}] \nonumber \\
        & ~~ + \lambda_e^{(N)} [\bar{v}_{70}~\bar{v}_{71}~\bar{v}_{72}~\bar{v}_{70}] + \mu_3^{(N)} [\bar{v}_{73}~\bar{v}_{71}~\bar{v}_{72}~\bar{v}_{73}] + \mu_2 [\bar{v}_{72}~\bar{v}_{71}~\bar{v}_{72}~\bar{v}_{72}] \\
        a\Bar{v}_8 & = u_s \Bar{\pi}_8 + \lambda p [\bar{v}_{70}~\bar{v}_{71}~0~\bar{v}_{73}] + \lambda_e^{(N)} [\bar{v}_{30}~\bar{v}_{31}~\bar{v}_{32}~\bar{v}_{30}] + \lambda_e^{(N)} [\bar{v}_{80}~\bar{v}_{81}~\bar{v}_{82}~\bar{v}_{80}]+ \mu_2 [\bar{v}_{82}~\bar{v}_{82}~\bar{v}_{82}~\bar{v}_{82}] \nonumber \\
        & ~~ + \mu_3^{(N)} [\bar{v}_{83}~\bar{v}_{81}~\bar{v}_{82}~\bar{v}_{83}] + \lambda p [\bar{v}_{80}~\bar{v}_{81}~0~\bar{v}_{83}]
    \end{align}
    \hrule
    \vspace{-6mm}
\end{subequations}
\end{small}
\end{figure*}
Then, using \eqref{cond_prob_eqn}, the steady-state conditional distribution vector satisfies the set of equations given in \eqref{eqn_v_s}. We resume the use of subscript $i$ notation and state the main result.
\vspace{-3mm}
\begin{theorem}
    Suppose the inter-arrivals at device $i$ are distributed as $\text{exp}(\lambda_i)$ and the service rates as $\text{exp}(\mu_{1i})$ and $\text{exp}(\mu_{2i})$. Let the service rate of the ES be distributed as $\text{exp}(\mu_3^{(N)})$. Then, the average AoI ${\Delta}^{(N)}_i(\bm{p},\bm{\mu}_1,\bm{\mu}_2)$ exists and is obtained by solving \eqref{steady_state_prob} and \eqref{eqn_v_s}. \hfill $\blacksquare$
\end{theorem}
\vspace{-3mm}
The proof of the above theorem follows by explicitly solving the set of linear equations \eqref{steady_state_prob} to get $\bar{\pi}_i$, substituting them in \eqref{eqn_v_s} and solving the latter set of equations.

\begin{figure*}[h]
\begin{small}
    \begin{subequations}
    \begin{align}\label{Avg_AOI_MF}
        {\Delta}_\phi(p_\phi,\mu_{1\phi},\mu_{2\phi},\rho) & = \frac{(1+\rho)}{\lambda_{1\phi}}\frac{\lambda_{1\phi}^3m_{1\phi}+\lambda_{1\phi}^2m_{2\phi} + \lambda_{1\phi}m_{3\phi}+\mu_{1\phi}\mu_{2\phi}(\mu_{1\phi}+\mu_{2\phi})(1+\rho)}{(\mu_{1\phi}+\lambda_{1\phi}p_\phi(1+\rho)+\mu_{1\phi}p_\phi\rho)(\mu_{2\phi}(\mu_{1\phi}+\mu_{2\phi})(1+\rho)+\lambda_{1\phi}\Bar{p}_\phi(\mu_{1\phi}+\mu_{2\phi}(1+\rho))} \\
        m_{1\phi}& := (1+\rho)p_\phi\Bar{p}_\phi,~ m_{2\phi}:= \mu_{2\phi}(1+\rho)+\mu_{1\phi}(1+(2-p_\phi)p_\phi\rho),~m_{3\phi}:= (1+\rho)(\mu_{1\phi}+\mu_{2\phi})^2-\mu_{1\phi}^2\bar{p}_{\phi}\rho
    \end{align}
    \end{subequations}
    \end{small}
    \hrule
    \vspace{-0.5cm}
    \end{figure*}
\vspace{-1mm}
% \begin{remark}
%     We note here that in the literature the offloading problem has been mostly approached using the framework of constrained optimal control rather than a game \cite{mao2016power,liu2021optimizing}. In this work, however, we model the problem using a non-cooperative game to account for the effect of other devices' policies on each device. Additionally, an AoI-based cost function helps us leverage its time responsiveness with respect to end-to-end packet delivery. We also note here that the employed static optimization framework is closer in spirit to policy optimization methods \cite{zhang2021multi} in reinforcement learning (RL) which have recently received great attention in a broad community. Thus, our work can be easily extended to the case with unknown ES parameters and packet arrival rates, which we leave as a promising future research direction.
% \end{remark}

\section{Mean-Field Game}
With the AoI calculations in the above subsection, we have provided a complete formulation of the $N$-user game problem. A suitable solution concept for the above game is that of seeking a Nash equilibrium policy \cite{bacsar1998dynamic}, i.e., a policy from which no user can deviate to receive a lower cost. However, its computation becomes intractable due to the high population regime. Thus, to alleviate this issue, we design Nash policies with the additional attractive feature that each user uses only its local policy information. In this regard, we leverage the framework of MFGs \cite{huang2007large}. Under the latter, we consider the limiting case ($N=\infty$) of the finite-user system (called the MF system). In this scenario, the individual user's deviations from equilibrium policies become insignificant due to the presence of infinite number of users. Hence, it suffices to consider the viewpoint of \textit{generic} user (representing the population of a specific type) which play against a mass distribution rather than each individual user. This then allows for the computation of the MF equilibrium in a completely decentralized manner, which constitutes an optimal policy of a generic user which is consistent with that of the population. With the above prelude, let us set up the MFG as follows.

Consider a generic device of type $\phi$ in the infinite population regime. The packets arrive at the device at mean rate $\lambda_\phi$. The tasks are split by employing an i.i.d.~Bernoulli distributed random variable with a mean $p_\phi$ into two independent Poisson processes with respective means $\lambda_\phi p_\phi$ and $\lambda_\phi \Bar{p}_\phi$. Further, the mean service rates of the generic transmitter and the generic local processor are given as $\mu_{1,\phi}$ and $\mu_{2,\phi}$, respectively. Let us define the average AoI of the packets of class 1 as ${\Delta}_\phi(p_\phi,\mu_{1\phi},\mu_{2\phi},\rho):= \lim_{N \rightarrow \infty}\Delta^{(N)}_i(\bm{p},\bm{\mu}_1,\bm{\mu}_2)$ where we define $\rho^{(N)}:= \frac{\lambda_e^{(N)}}{\mu^{(N)}_3}$, and $\rho:= \lim_{N \rightarrow \infty}\rho^{(N)}$. The latter exists, for instance, when the service rate of the ES increases proportionally to the number of devices, i.e., when $\mu_3^{(N)} = N\mu_3$ for $\mu_3>0$, which is also what we consider for numerical evaluation purposes. The term $\rho$ serves as the MF approximation to the coupling term $\rho^{(N)}$ in the finite-device system and can be viewed as the mean load on the ES in the infinite-device system. Then, we have the generic device optimization problem as follows.
\vspace{-2mm}
\begin{problem}[Generic device optimization problem]\label{problem:generic_user}
    \begin{align}
        \min_{(p_\phi,\mu_{1\phi},\mu_{2\phi}) \in [0,1] \times \mathbb{R}^2} &  J_{\rho}(p_\phi,\mu_{1\phi},\mu_{2\phi}) \nonumber \\
        \mbox{s.t.} \quad &  \mu_{1\phi} \leq P_{\phi,max} \nonumber \\
        & \mu_{2\phi} \leq f_{\phi,max}
        \end{align} 
    where $J_{\rho}(p_\phi,\mu_{1\phi},\mu_{2\phi}) \!:=\! t_{L_\phi}\mu_{1\phi} + t_{T_\phi}\eta \mu_{2\phi}^3 \!+\! V {\Delta}_\phi(p_\phi,\mu_{1\phi},\mu_{2\phi},\rho)$,
${\Delta}_\phi(p_\phi,\mu_{1\phi},\mu_{2\phi},\rho)$ is given in \eqref{Avg_AOI_MF} and $t_{L_\phi},t_{T_\phi}$ denote the busy periods of $L_\phi$ and $T_\phi$, respectively, of the generic device.
\end{problem}

Consequently, the MFG is defined using the optimality and the consistency conditions as follows:
\begin{enumerate}
    \item Optimality: $(\hat{p}_\phi,\hat{\mu}_{1\phi},\hat{\mu}_{2\phi}) \!=\! \arg\min J_\rho(p_\phi,\mu_{1\phi},\mu_{2\phi})$,
    \item Consistency: $\hat{\rho} = \frac{1}{\mu_3}\mathbb{E}_{\phi}\left[\frac{\lambda_\phi \hat{\Bar{p}}_\phi \hat{\mu}_{2\phi}}{\lambda_\phi \hat{\Bar{p}}_\phi + \hat{\mu}_{2\phi}}\right]$.
\end{enumerate}

Briefly, for a given value of $\rho$, the generic user solves for an optimal policy using the optimality condition. Consequently, it uses the obtained policy to regenerate $\rho$ using the consistency condition. The mean-field equilibrium (MFE), which constitutes the pair $((p_{\phi, \text{MFE}}, \mu_{1\phi, \text{MFE}}, \mu_{2\phi,\text{MFE}}), \rho_{\text{MFE}})$, is then given by the fixed point of the composite map induced by 1) and 2) for all $\phi \in \Phi$. Detailed fixed point iteration process is given in Algorithm~\ref{alg1} below.

\begin{algorithm}[h!]
	\caption{Fixed point iteration for a generic device}
	\begin{algorithmic}[1] \label{alg:FPI}
        \STATE {\textbf{Input:} $V,\eta,\mu_3,\lambda_\phi, ~\forall \phi$ \hfill\# system parameters}
        \STATE {\textbf{Input:} $\epsilon$ \hfill \# approximation parameter}
        \STATE {\textbf{Input:} $\gamma$ \hfill \# iteration step size}
		\STATE {Initialize: $\hat{\rho}, p_\phi^{(0)},\mu_{1\phi}^{(0)},\mu_{2\phi}^{(0)}=0,~\forall \phi$}
        \WHILE{$|\hat{\rho}^{(m)} - \hat{\rho}^{(m-1)}| < \epsilon$}
        \STATE $(\hat{p}^{(k)}_\phi, \hat{\mu}_{1\phi}^{(k)},\hat{\mu}_{2\phi}^{(k)})\leftarrow \text{argmin } J_{\hat{\rho}^{(k-1)}}(p_\phi^{(k-1}\!\!,\mu_{1\phi}^{(k-1)}\!\!,\mu_{2\phi}^{(k-1)})$
        \STATE $\hat{\rho}^{(k)} \leftarrow (1-\gamma)\hat{\rho}^{(k-1)} + \gamma \mathbb{E}_\phi\frac{\lambda_\phi(1-\hat{p}^{(k)}_\phi)\hat{\mu}_{2,\phi}^{(k)}}{\mu_3(\lambda_\phi(1-\hat{p}^{(k)}_\phi)+\hat{\mu}_{2\phi)}^{(k)}}$
        \ENDWHILE
        \STATE \textbf{Output: } Last iterate: $\hat{\rho}^{(m)}, (\hat{p}^{(m)}_\phi, \hat{\mu}_{1\phi}^{(m)},\hat{\mu}_{2\phi}^{(m)}), ~\forall \phi.$
	\end{algorithmic}
    \label{alg1}
\end{algorithm}
\vspace{-3mm}
\section{Numerical Results}
\begin{figure}[t]
    \centerline{\includegraphics[width=0.85\columnwidth]{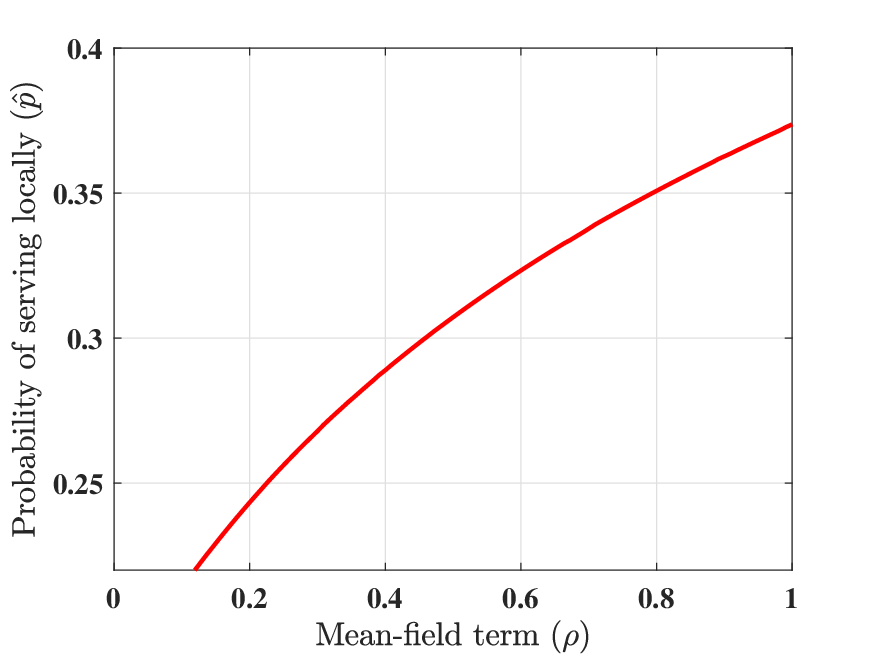}}
    \caption{\small{The optimal probability $\hat{p}$ as a function of the MF term $\rho$.}}
    \label{Fig:sim1}
    \vspace{-0.4cm}
\end{figure}

\begin{figure}[t]	
    \centerline{\includegraphics[width=0.9\columnwidth]{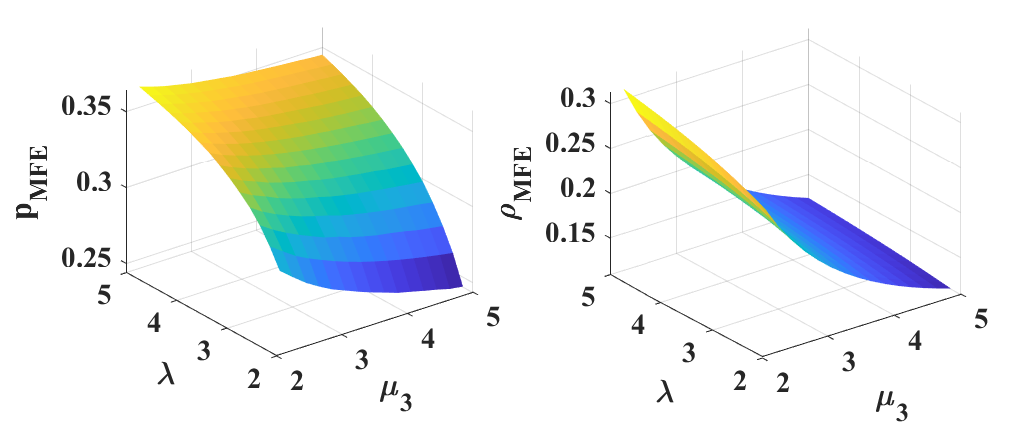}}
	\vspace{-0.3cm}
	\caption{\small{The equilibrium probability $p_{\text{MFE}}$ and equilibrium load $\rho_{\text{MFE}}$ with generic arrival rate $\lambda$ and $\mu_3$.}}
	\label{Fig:sim2}
    \vspace{-0.4cm}
\end{figure}

Here, we provide a numerical computation of the MFE for a population of a single type. In the first numerical study, in Fig.~\ref{Fig:sim1}, (for $V=10,\eta=5,\lambda=2.5,P_{T,max}=1,f_{max}=0.3$) we observe that as the mean loading at the ES increases (on the $x$-axis), the optimal probability of using the local processor (on the $y$-axis) increases and that of offloading to the ES decreases. This should be expected since if the ES is heavily loaded, the device is better-off serving tasks locally to incur a lower AoI. Next, in Fig.~\ref{Fig:sim2}, (for $V=10,\eta=0.5,P_{T,max}=1,f_{max}=0.3$) we plot the variation of the MFE as a function of the arrival rate $\lambda$ and service rate $\mu_3$. From the figure, we observe that at equilibrium increasing arrival rate offloads more computations, thereby increasing the ES loading. On the other hand, increasing ES service rate increases offloading by the device, but with a decrease in the mean ES loading, thereby suggesting a slower than linear optimal rate of task offloading by the devices.

\vspace{-1mm}
\section{Discussion and Conclusion}
As a recap, in this work we have considered a timely task computation problem in a dense-user MEC system where the devices can either process their tasks on their local processors or offload them to an ES. We have developed a finite-user Nash game and a MFG model for the task offloading problem in MEC systems, and provided a low complexity algorithm to compute decentralized equilibrium solutions to the MF system. In the future, we plan to investigate the theoretical aspects of the above developments, particularly related to the characterization of the conditions ensuring existence and (possible) uniqueness of the MFE in ultra-dense user networks.

\bibliographystyle{unsrt}
\bibliography{IEEEabrv,references} 

\end{document}